\documentclass[runningheads]{llncs}

\usepackage{graphicx}
\usepackage{natbib}
\usepackage{array}
\usepackage{hyperref}
\usepackage{amsmath}
\usepackage{tikz}

\newcommand{\tone}{44.9$^{+13.8}_{-12.7}$\,\,}
\newcommand{\ttwo}{30.9$^{+15.1}_{-12.2}$\,\,}

\newcommand\pasp{PASP} 
\newcommand\apjs{ApJS}  
\newcommand\apj{ApJ}    
\newcommand\aap{A\&A}  
\newcommand\araa{ARA\&A}  
\newcommand\aaps{A\&AS}
\newcommand\aj{AJ}  

\begin{document}
\title{Medium-band photometric reverberation mapping of AGNs at $0.1 < z < 0.8$}
\subtitle{Techniques and sample}

\author{
        E.\,Malygin \inst{1} 
      \and 
        R.\,Uklein \inst{2}   
      \and 
       E.\,Shablovinskaya \inst{2}
       \and
      A. Grokhovskaya \inst{2}
       \and
        A. Perepelitsyn \inst{2}
       }

\institute{
           Kazan Federal University, Kazan, Russia, \email{playground@mail.ru}
         \and 
          Special Astrophysical Observatory RAS, Nizhny Arkhyz, Russia, \email{uklein.r@gmail.com}
          }
\authorrunning{Malygin et al.}

\maketitle              
\begin{abstract}
 The most popular method of the broad-line region size estimation in active galactic nuclei (AGN) is the reverberation mapping based on measuring the time delay between the continuum flux and the flux in the emission lines. In our work, we apply the method of photometric reverberation mapping in mid-band filters, adapted for observations on the 1-m Zeiss-1000 telescope {of} Special Astrophysical Observatory of Russian Academy of Sciences, for the study of AGN with broad lines in the range of redshifts $0.1 < z < 0.8$. This paper provides a sample of 8 objects, describes the technique of observations and data processing for 2 studied objects to demonstrate the stability of the used method.
\end{abstract}

%
\section{Introduction}
\label{intr}
\begin{figure}
    \centering
    \includegraphics[width=0.9\textwidth]{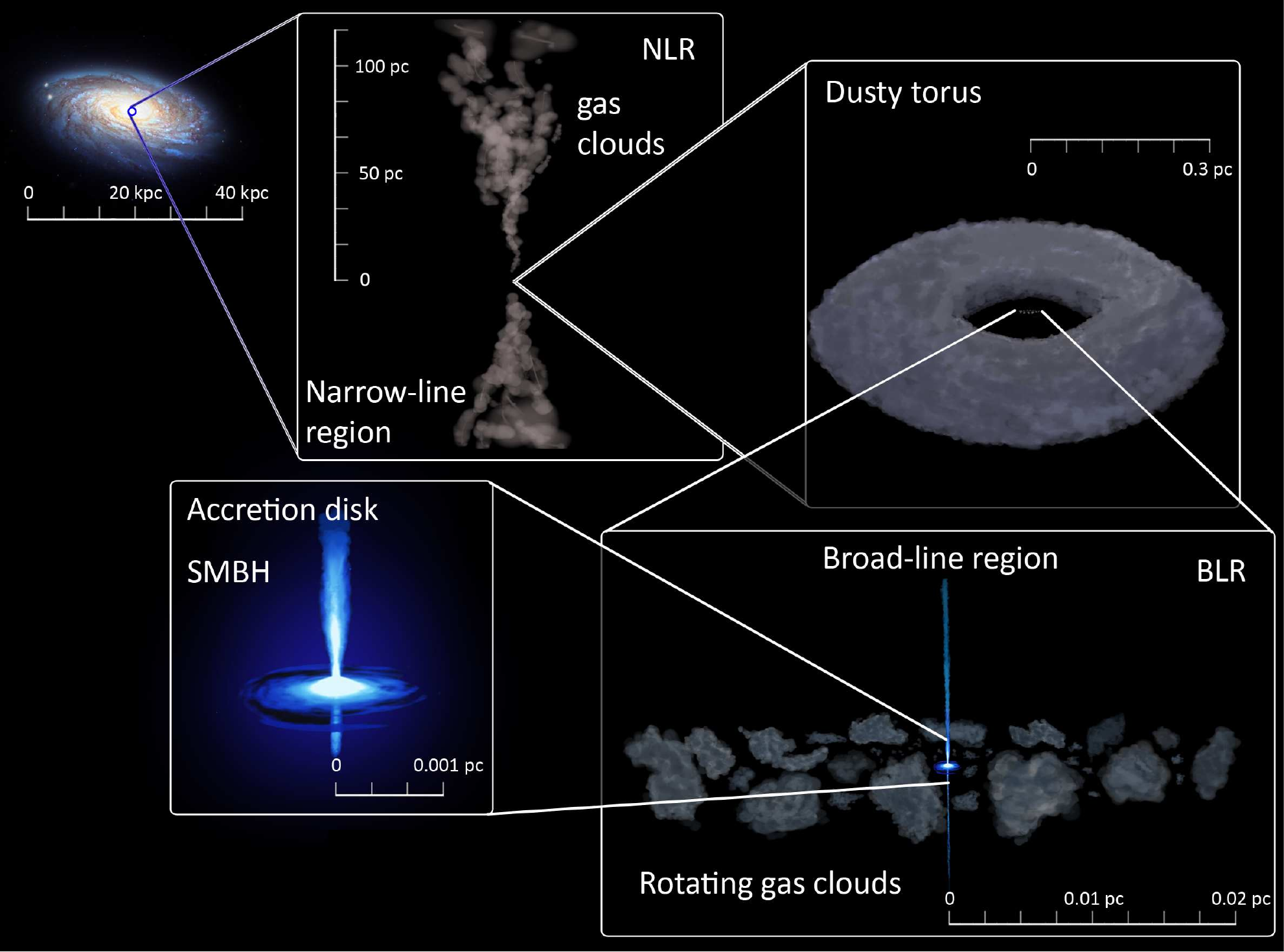}
    \caption{The illustration of modern conceptions of AGN structure.}
    \label{AGN}
\end{figure}

Active galactic nuclei (AGN) are bright compact areas, emitting up to 90\% of the energy of the entire galaxy. Because of the small size of the central object and the high gas velocity, it is now assumed that the supermassive black hole (SMBH) is present in the center of the galactic nucleus, and the enormous luminosity is provided by the processes of the matter accretion into the SMBH. The observed energy distribution, as well as the AGN spectrum, indicates that the nucleus is a multicomponent system, which observational properties depend on the orientation relative to the observer. This is known as the Unified model  \citep{Ant93, UnMod}, which is illustrated in Fig. \ref{AGN}. According to this model, the SMBH is surrounded by a hot disk of accreting matter ($\sim$0.001 pc from the central source), which hard radiation excites the gas in the broad-line region (BLR) at scales of 0.01-0.1 pc and is re-emitted in emission lines broadened due to the high gas velocity (up to $10^4$ km/s). 

The reverberation mapping method \citep[RM,][]{BlanfordMKee82} consists of measuring the time lag $\tau$ between the continuum radiation of the accretion disk and the radiation in the emission lines produced in the BLR in AGN. The BLR size is assumed to be $R_{\rm BLR} \equiv c\tau$, where $c$ is the speed of light \citep{Peterson93}. Then, the $R_{\rm BLR}$ can be determined by measuring the time lag $\tau$ between the continuum and line light curves.
According to the virial ratio the SMBH mass is related to the BLR size $c\tau$ and the gas velocity in the BLR $\upsilon_{\rm line}$ as: $$M_{\rm SMBH}=fc\tau\upsilon_{\rm line}^2G^{-1},$$
where $G$ is the gravitational constant and $f$ is a dimensionless factor of the order of one depending on the BLR structure and kinematics and the inclination of the system relative to the observer. Thus, measuring the time delays provides one with the SMBH mass estimation. Yet, the RM method requires the accumulation of a long series of observational data that makes harder its wide application.  That is the reason why presently this method was applied only for approximately 100 nearest quasars \citep[e.g.][]{Sh09, BentzKatz15, Du16,Jiang16,Nunez17, I17}. 

However, to trace the evolution of the SMBH masses the extension of the AGN sample with known sizes $R_{\rm BLR}$ to more distant redshifts is needed. For AGN the linear relation between the BLR size and AGN luminosity was found: $R_{\rm BLR} \propto L^{\alpha}$, where $L$ could be measured in different spectral bands such as 5100\AA\ \citep{Kaspi05, bentz09} or definite emission lines \citep[e.g. MgII in][and references therein]{Czerny19}.

Our investigation is dedicated to complement the existing relation of $R_{\rm BLR}(L)$ by new measurements of $R_{\rm BLR}$ = $c\tau$ for the distant AGNs up to $z \sim 0.8$ using a sample of objects that have not been studied in other reverberation {campaigns}. Besides, we adopt the photometric RM method \citep{Haas11} to mid-band observation with the 1-meter class telescope. 

The paper is formed as following. In Sec. 2 the sample and the methods of observations and data reduction are given. Sec. 3 contains the first results of the applied observational technique with the description of the local standards and the preliminary light curves. The conclusions and main ideas are shortly given in Sec. 4. 

\section{Observations}
\subsection{Sample}

Our sample consists of 8 AGNs with broad lines (equivalent width $W_{\lambda} > 200$\AA) in the range the redshifts $0.1 < z < 0.8$ with the brightness limited to $m <$ 20 mag. The sample includes only near-polar objects (Dec $ > 68^{\circ}$), thus we were able  to observe them throughout the whole year. The final sample is shown in Table \ref{T:sample}. 

\begin{table*}[]
\centering
\footnotesize
\begin{tabular}{|c|l|m{1.8cm}|c|c|c|m{1.05cm}|}
\hline
 & Name			& RA Dec & Mag  & $z$  & $\tau$  & Filters\\
(\#) & (1)  & (2)  & (3) & (4) & (5) & (6)  \\
\hline
1 & 2MASX J08535955+7700543		& 08$^{\rm h}$53$^{\rm m}$59$^{\rm s}$.4 $+77^\circ00'55''$ & $17.0$ & 0.106  & $27$ & SED725 SED700 \\
\hline
2 & VII Zw 244 			        & 08$^{\rm h}$44$^{\rm m}$45$^{\rm s}$.3 $+76^\circ53'09''$ & $15.7$ & 0.131  & $34$ & SED550 SED525 \\
\hline
3 & SDSS J093702.85+682408.3	& 09$^{\rm h}$37$^{\rm m}$02$^{\rm s}$.9 $+68^\circ24'08''$ & $18.0$ & 0.294  & $47$ & SED625 SED600 \\
\hline
4 & SDSS J094053.77+681550.3	& 09$^{\rm h}$40$^{\rm m}$53$^{\rm s}$.8 $+68^\circ15'50''$ & $19.4$ & 0.371  & $59$ & SED900 SED875 \\
\hline
5 & SDSS J100057.50+684231.0 	& 10$^{\rm h}$00$^{\rm m}$57$^{\rm s}$.5 $+68^\circ42'31''$ & $19.0$ & 0.499  & $80$ & SED725 SED700 \\
\hline
6 & 2MASS J01373678+8524106 	& 01$^{\rm h}$37$^{\rm m}$36$^{\rm s}$.7 $+85^\circ24'11''$ & $16.6$ & 0.499  & $79$ & SED725 SED700 \\
\hline
7 & SDSS J095814.46+684704.8 	& 09$^{\rm h}$58$^{\rm m}$14$^{\rm s}$.4 $+68^\circ47'05''$ & $19.7$ & 0.662  & $92$ & SED800 SED775 \\
\hline
8 & GALEX 2486024515200490156 	& 10$^{\rm h}$01$^{\rm m}$51$^{\rm s}$.6 $+69^\circ35'27''$ & $19.6$ & 0.847  & $124$ & SED900 SED875 \\
\hline
\end{tabular}
\caption{Observed sample of AGN: (\#) identification number in the sample; (1) galaxy name; (2) coordinates for the J2000 epoch; (3) magnitude in $V$ band; (4) redshift $z$; (5) roughly estimated time delay $\tau$ in days from the continuum or emission line luminosity (see text for details); (6) filters used to measure line and continuum fluxes.}
\label{T:sample}
\end{table*}

\begin{figure*}[tp!] 
\centering
\footnotesize
\includegraphics[width=0.9\linewidth]{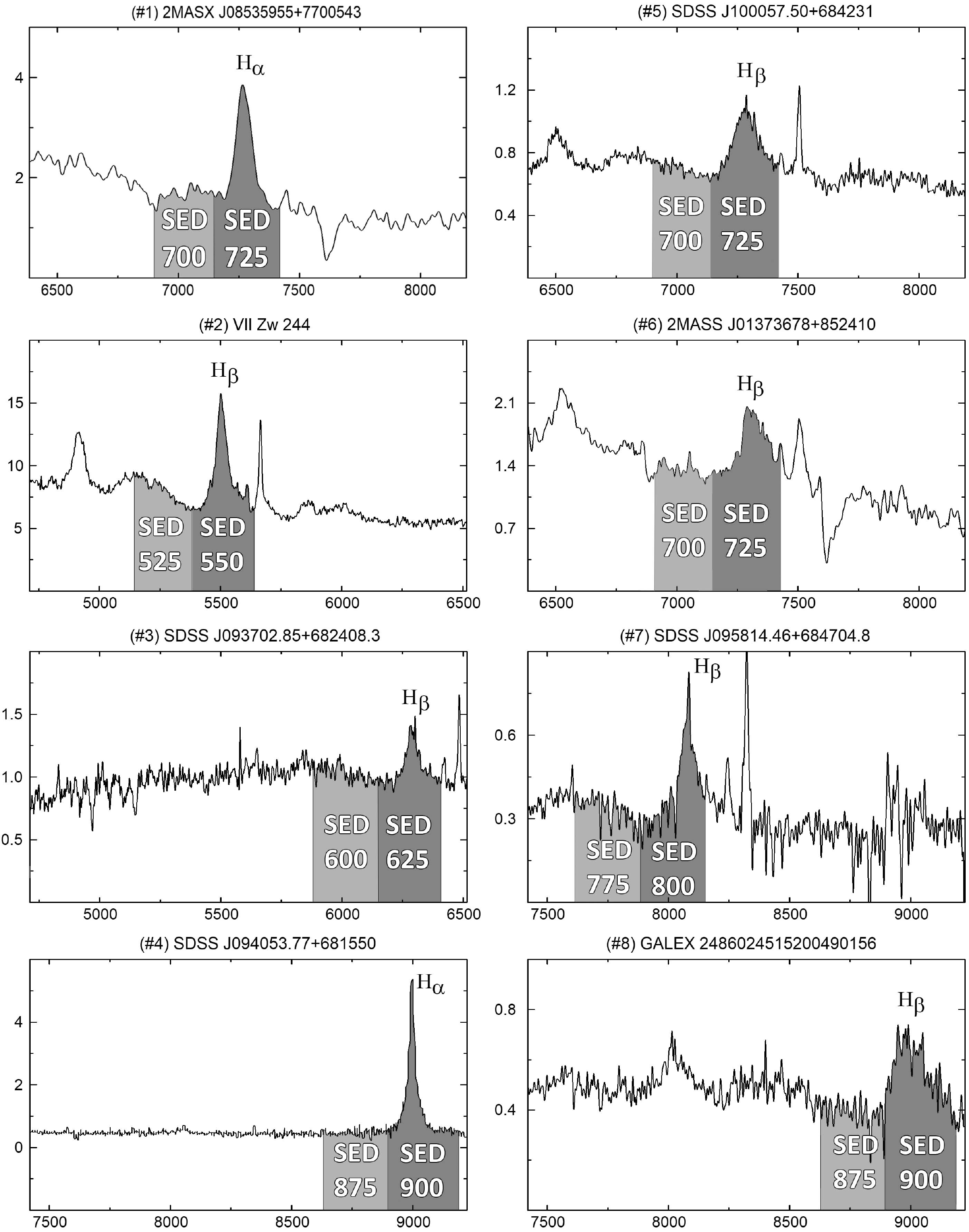}
\caption{SED filters bands overplotted on the spectra of the studied AGN. Flux $F_{\lambda}$ in units of $10^{-16}$ erg/cm$^{2}$/s/\AA\ depending on the wavelength in \AA\ given in the observer rest frame. The spectra of \#1 and \#6 are obtained from \cite{spec2}, \#2 is from \cite{spec1} and the others are from \cite{spec}.}
\label{F:filters}
\end{figure*}

Each object is observed in two filters  selected so that they cover the broad emission line H$_{\beta(\alpha)}$ and the near-by continuum, which was used to subtract the contribution of the variable continuum from the emission line. The experiment uses medium-band interference filters\footnote{Edmund Optics, \url{https://www.edmundoptics.com}. \\ For the main parameters of middle-band SED filters see:  \url{https://www.sao.ru/hq/lsfvo/devices/scorpio-2/filters_eng.html}.} with the spectral energy distribution (SED) of a 250\AA\ bandwidth, covering the 5000--9000\AA\, range with 250\AA-step. The selection of filters with their bandwidth overplotted on the spectra of  the studied objects is illustrated in Fig.~{\ref{F:filters}}.  The spectra are taken from \cite{spec}, \cite{spec1}, and \cite{spec2}.

From the known radius-luminosity relation $R_{\rm BLR} - L$ for the H$_{\beta}$ line the expected time delays $\tau$ were calculated for the sample (see Table~{\ref{T:sample}}, column 5) in the following way. For the objects with redshifts up to 0.5 - objects \#1,3-5 - the flux $F_{\lambda}$ at 5100\AA\ was integrated in the range 4400-5850\AA\ in the rest frame from the spectra obtained from the literature; the flux $F_{\lambda}$ was divided by the window-width in order to get the monochromatic luminosity. The 4400-5850\AA\ range contains some strong emission lines as variable H$_\beta$ and approximately constant O[III] but as the integrated band is wide we assume that the lines contribution is not crucial especially for rough estimation of the time delays. To convert the flux to the luminosity $L_{\lambda}$ at 5100 \AA\, it was multiplied with $4\pi D^2$, where $D$ is the AGN luminosity distance calculated using the cosmological parameters $\Omega_0=0.3036$ and $H_0=68.14$ km/s/Mpc. Note here that the contribution of the host galaxy was not considered.  Then we applied the relation $R_{\rm BLR}-L_{\lambda}({5100})$: 
$$
\lg(R_{\rm BLR}) = -21.3^{+2.9}_{-2.8} + 0.519^{+0.063}_{-0.066} \lg(\lambda L_{\lambda}), 
$$
where $ L_{\lambda}=L_{\lambda}({5100})$ is a luminosity at 5100\AA, and $R_{\rm BLR}$ is the BLR size in the H${_\beta}$ line  \citep{RL5100}.

In the case of $z > 0.5$, as well as for the object \#2, which spectral data used by us in calculations are available only in a small wavelength range (4075-5883\AA), the $L_{\lambda}({5100})$ range goes beyond the available optical spectra. In this regard, for objects \#2,6-8 we used the relation with the line $L_{\lambda}({\rm H}_{\beta})$ from \cite{Greene10}:
$$
\lg(R_{10}) = 0.85 \pm 0.05 + (0.53\pm0.04) \lg[L_{43}({\rm H}_{\beta})], 
$$
where $R_{10}=R_{\rm BLR}/10$ lt days is the size of the BLR region, normalized to 10 lt days, $L_{43}({\rm H}_{\beta}) = L_{\lambda}({\rm H}_{\beta})/10^{43}$ erg/s is the luminosity in the H$_{\beta}$  line, normalized to $10^{43}$ erg/s. In Table  \ref{T:sample} the rough estimate of the expected time delays $\tau$ are given with an accuracy of 10\%.

\subsection{Observational technique and reduction }

Since February 2018, observations of the AGN sample have been carried out monthly on gray and bright nights at 1-m Zeiss-1000 telescope of the Special Astrophysical Observatory of Russian Academy of Sciences (SAO RAS) using MaNGaL \citep[MApper of Narrow GAlaxy Lines,][]{Pulkovo} and MMPP (Multi-Mode Photometer-Polarimeter) devices in photometric mode with 10 medium-band interference SED filters. The size of the field of view (FoV) was 8.$'7\times 8.'7$ for MaNGaL and 7.$'2\times 7.'$2 for MMPP. Up to now, around 15 epochs  were acquired on average for majority of the objects in the sample.

Three different detectors were used during the observations: Andor CCD iKon-M 934 (1024$\times$1024 px), Andor Neo sCMOS (2560$\times$2160 px) and Raptor Photonics Eagle V CCD (2048$\times$2048 px). The quantum efficiency of these receivers in the needed bands is shown in Table~\ref{T:qe}. Water cooling was used for all three cases to minimize noise.

\begin{table}[]
\centering
\footnotesize
\begin{tabular}{lccccc}
Detector  & 5500\AA\ & 6000\AA\ & 7000\AA\ & 8000\AA\ & 9000\AA\  \\
\hline
Andor iKon-M 934    & 95\%      & 96\%      & 91\%      & 77\%      & 47\%       \\
Andor Neo sCMOS     & 54\%      & 56\%      & 49\%      & 31\%      & 14\%       \\
Raptor Eagle V CCD         & 92\%      & 95\%      & 89\%      & 75\%      & 50\%       \\
\hline
\end{tabular}
\caption{Quantum efficiency of detectors in the studied photometric bands.}
\label{T:qe}
\end{table}

Observations of the sample were alternated with observations of spectrophotometric standard stars provided by \citet{Oke90}. The standards were observed almost simultaneously with the object field in the same filter to minimize the variations of atmospheric transmission; moreover, even observing the standards at the close zenith distance we tried to obtain the extinction coefficient every night and take it into account. Such strategy was used to create a list of comparison stars  with known AB-magnitudes in the FoV of the sample. 

During each observational night, we received calibration images (flat frames at the twilight sky for each filter, bias/dark) to correct data for additive and multiplicative errors. For each object the series of images (3 and more) were taken, the exposure times depend on the object brightness, weather conditions and are ranged from 2 to 10 minutes. For correct statistics each frame is processed independently, and statistical evaluation is made by averaging the random value by robust methods giving its unbiased estimate. In this case, the photometric errors are the rms of the robust distribution.

The method of aperture photometry was used to determine the absolute flux of the objects in instrumental units. Therefore, to correctly estimate the sky background, the traces of cosmic rays which were close to the object were removed from images. 

For better photometric accuracy needed for the variability studies of AGN, the AB fluxes were first calculated using the spectrophotometric standards, and further step was to re-calibrate them with the differential photometry using the local standards from the object image FoV.

\section{Results}
\subsection{Local standards}

\begin{figure}
    \centering
    \includegraphics[width=0.42\textwidth]{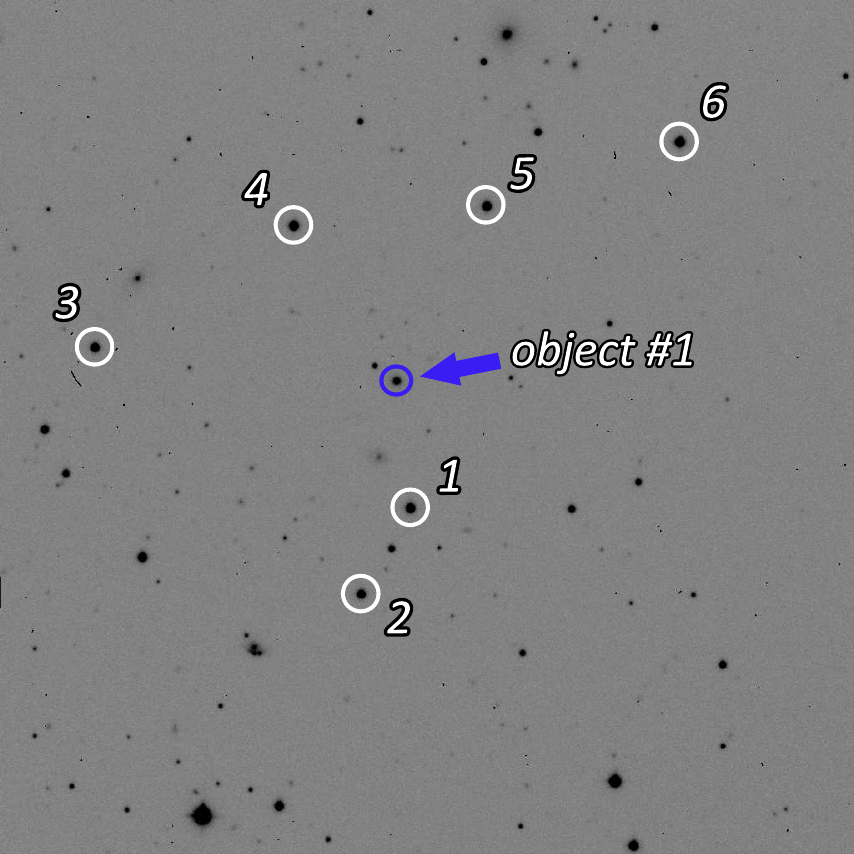}
    \includegraphics[width=0.42\textwidth]{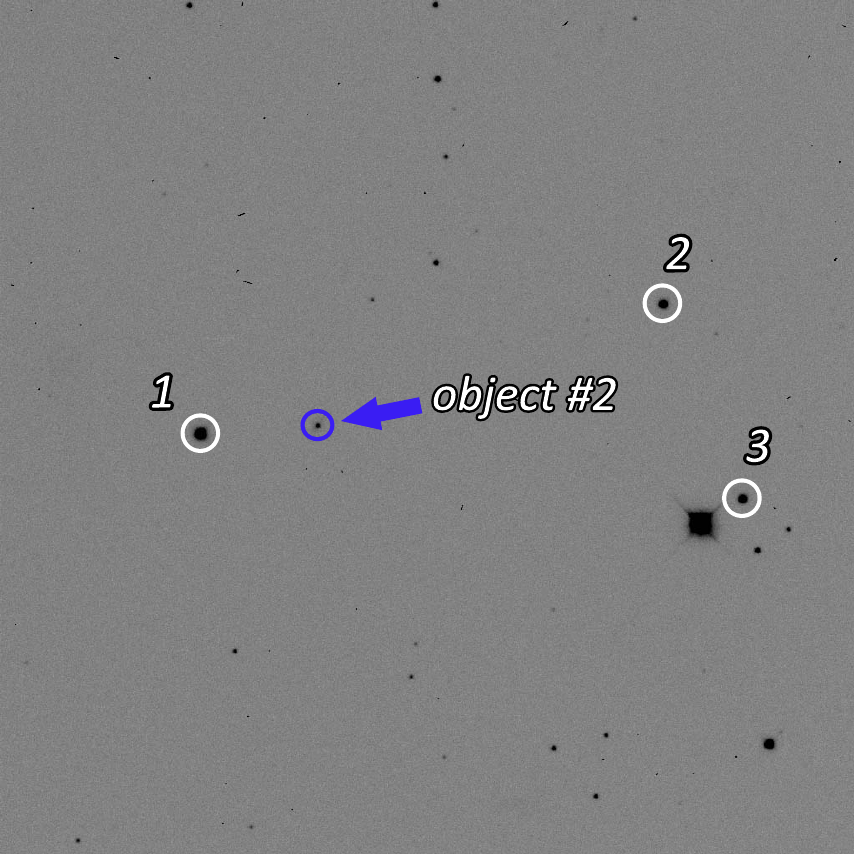}
    \caption{Direct images of AGNs: the object \#1 obtained in the SED725 filter with 6 local standards candidates marked (\textit{left}) and object \#2 obtained in the SED550 filter by CCD Eagle V [MMPP] with 3 local standards candidates marked (\textit{right}).}
    \label{field}
\end{figure}

\begin{figure}
    \centering
    \includegraphics[width=0.48\textwidth]{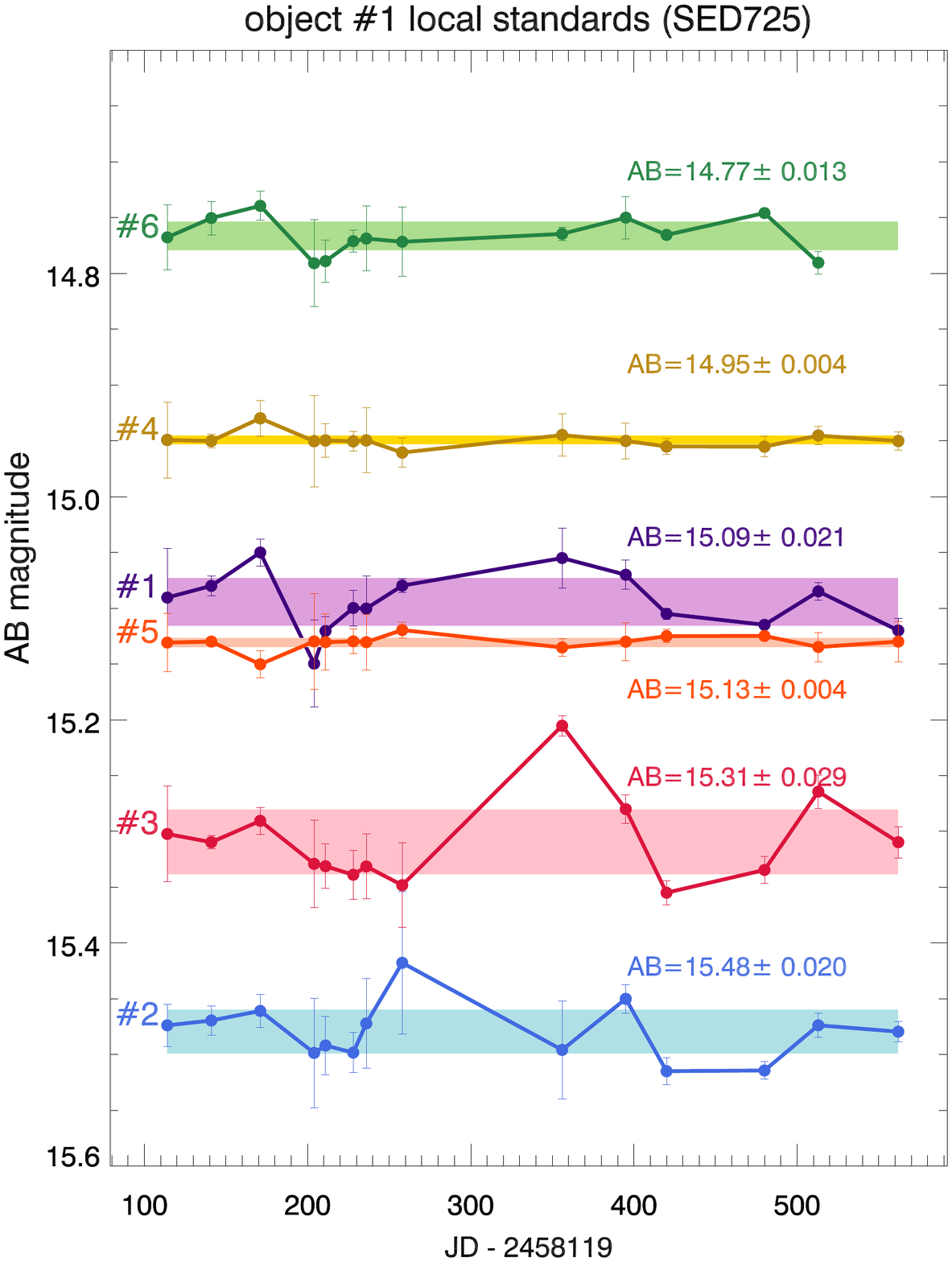}
    \includegraphics[width=0.48\textwidth]{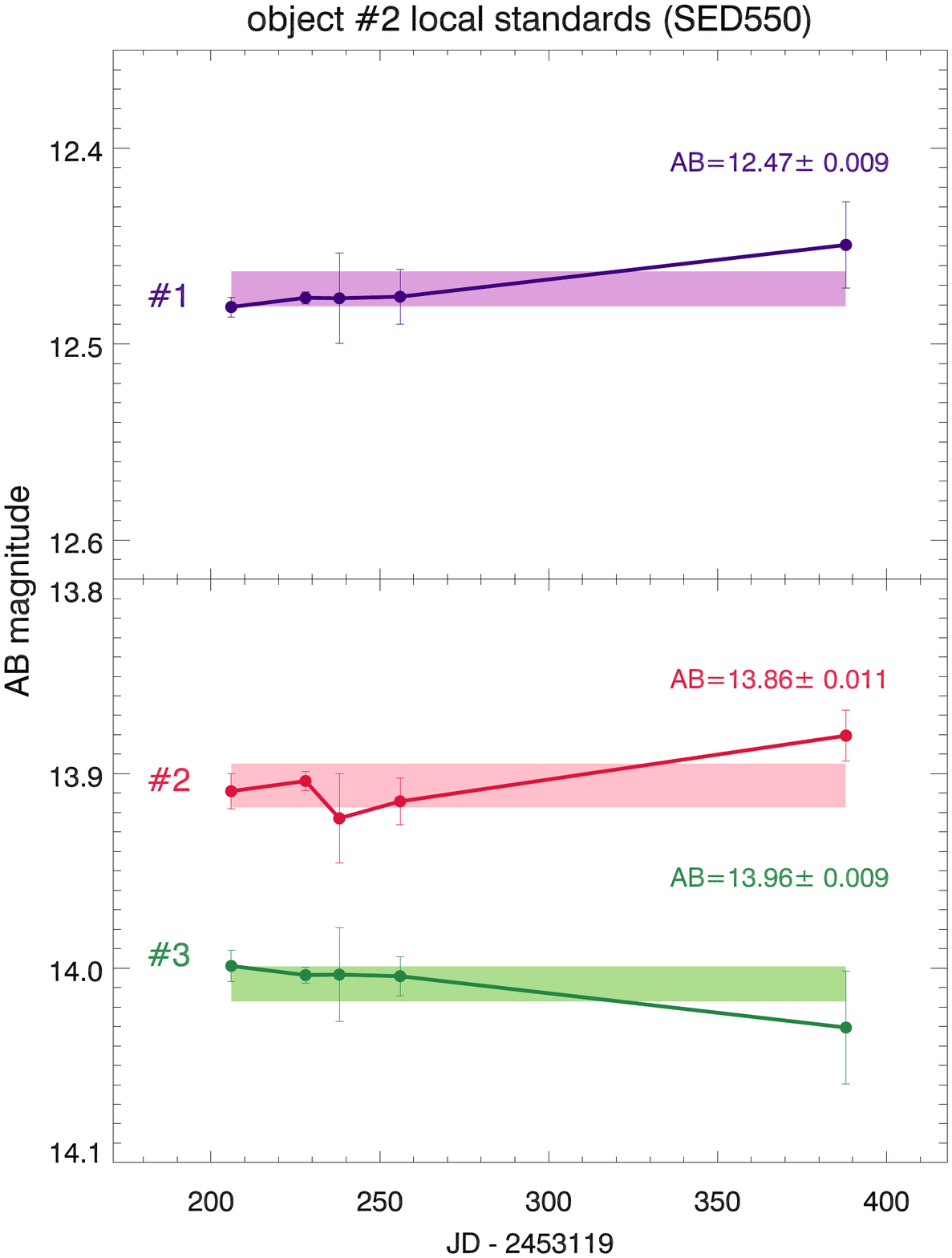}
    \caption{The light curves of the local standards candidates: 6 stars for object \#1 in SED725 (\textit{left}) and 3 stars for object \#2 in SED550 filter (\textit{right}).}
    \label{stds}
\end{figure}

The main methodological result obtained during the first year of observation{s} is the definition of local standards in the field of each sample objects. In Fig. \ref{field} we show the field of the object \#1 (2MASX J08535955+7700543) obtained in the SED725 filter and object \#2 (VII Zw 244) obtained in the SED550 filter by Raptor Eagle V CCD (MMPP) with marked local standards. Within the stable photometric nights, we bind the field with known spectrophotometric standards. To examine the variability of the stars in the fields, the obtained fluxes were normalized to the most frequently observed and the most stable star flux. Thus, the resulting fluxes of stars do not depend on weather conditions on the selected night. Fig. \ref{stds} presents the normalized light curves of local standards in objects \#1 and \#2 fields in AB-magnitudes. In Fig. \ref{stds} the regions corresponding to the 3$\sigma$ confidence area are also shown. For differential photometry, the stars with the smallest scatter were chosen. It is also seen that the average error of the absolute binding was of the order of 0.03 mag.

Once local standards have been defined, the flux of the object is measured relatively to them, minimizing errors introduced by variations in the atmosphere.

\subsection{First light of light curves}

\begin{figure}
    \centering
    \includegraphics[width=0.48\textwidth]{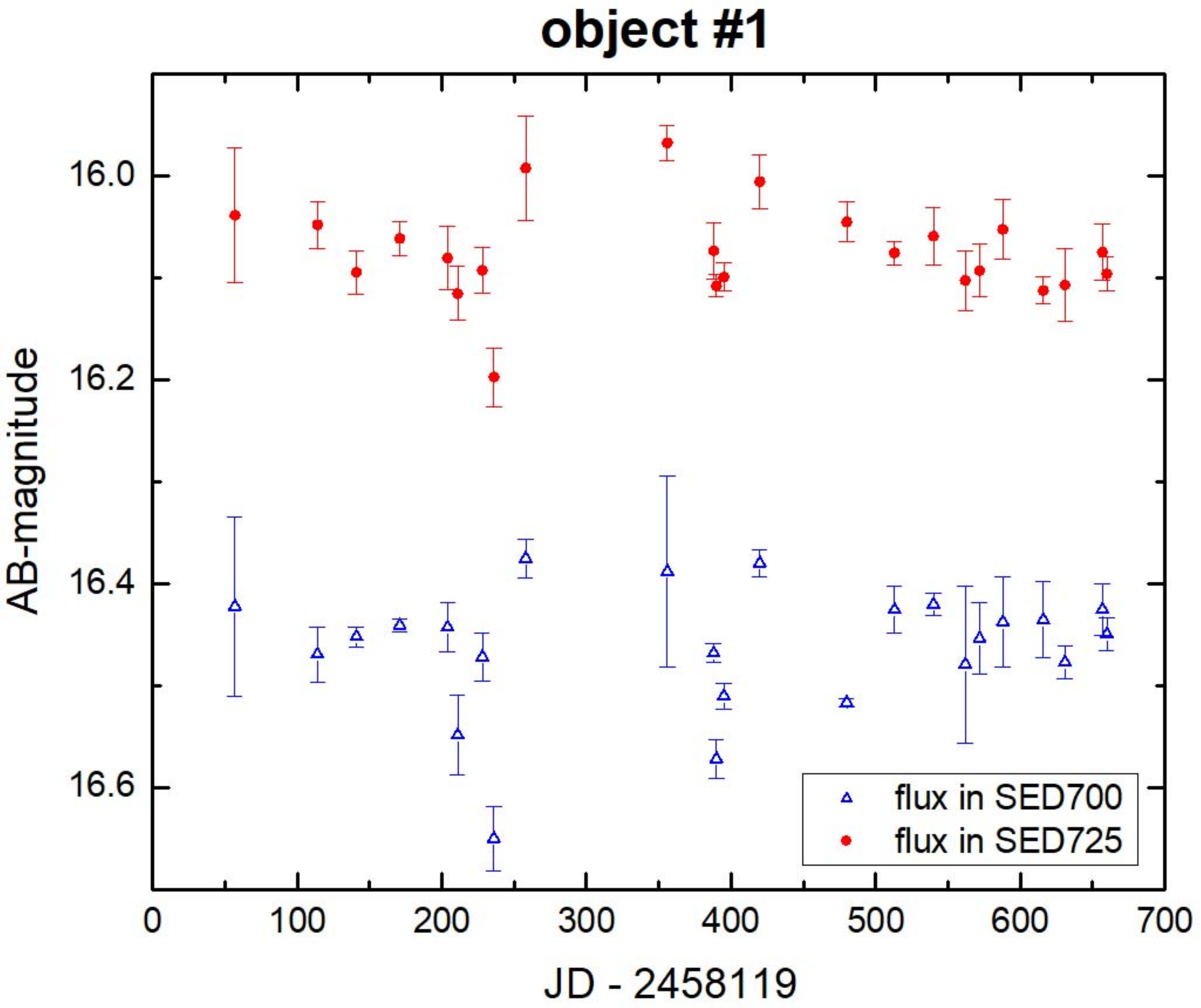}
    \includegraphics[width=0.48\textwidth]{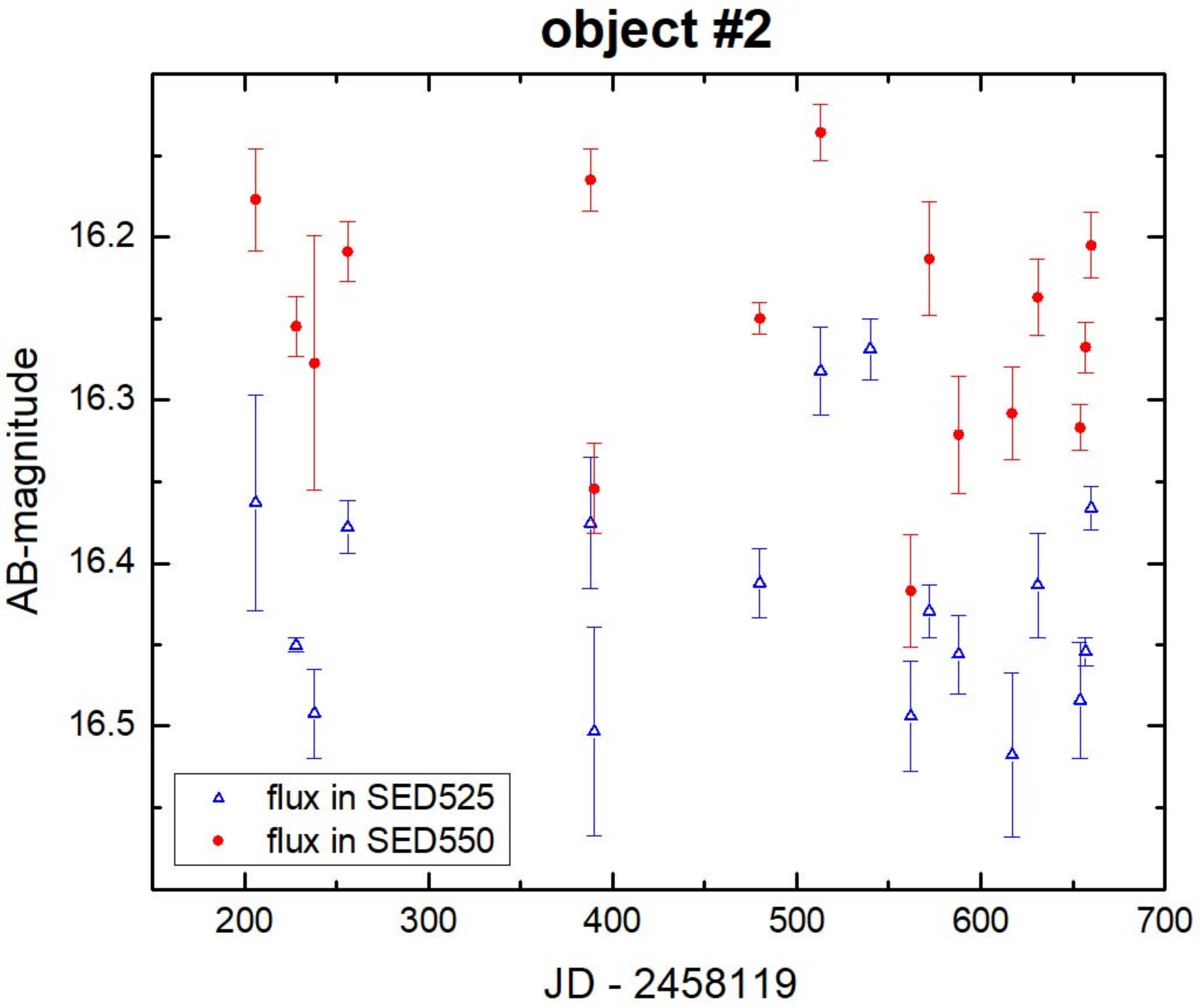}
    \caption{The light curves of object \#1 ({\it left}) and object \#2 ({\it right}) in the line (circles) and near-by continuum (triangles). The Julian dates starts on January 1, 2018.}
    \label{F:LC}
\end{figure}

The fluxes of the studied AGN were carried out relatively to the most stable reference stars --- local standards. The light curves in the continuum and the line for one of the most intensively observed AGN, object \#1 are presented in Fig.~{\ref{F:LC}}, left. Also, in Fig. \ref{F:LC}, right, the same light curves for object \#2 are shown.

To investigate the time lag between the light curves we have used the {\tt JAVELIN} code \citep{zu16,jav}. The first step is to build a continuum model to determine the dumped random walk (DRW) parameters of the continuum light curve. As a result, the posterior distribution of the two DRW parameters (amplitude and timescale) of the continuum variability is calculated by {\tt JAVELIN} using Markov chain Monte Carlo (MCMC). The second step is to interpolate the continuum light curve based on the posteriors derived on the first step, and then to shift, smooth, and scale each continuum light curve to compare with the observed line light curve. Finally, it derives the posterior distribution of the time lag and other parameters. 

Fig.~\ref{ccf} shows the posterior distribution of {\tt JAVELIN} time delays for objects \#1 and \#2. This is a distribution of the number of hits N during the MCMC sampling, which sum across all bins gives the sample size. We used at least 10,000 samples. Solid red line marks position of the local peak median, whereas dashed blue line marks expected lag from calibration relations (Table \ref{T:sample}, column 5). Our preliminary lags are \tone and \ttwo days for objects \#1 and \#2 respectively. Low and high estimates of the lags correspond to the highest posterior density interval calculated by {\tt JAVELIN}.


\subsubsection{JAVELIN analysis for object \#1}
The {\tt JAVELIN} method involves sampling using the DRW, which ultimately gives us slightly different results from one chain to another. In general, the overall picture for the data set can be traced, however, we can choose the clearest histograms for analysis.\\
In case of the object \#1, we see a stable result for 24 epochs (Fig.~\ref{ccf}, left panel): 
\begin{enumerate}
\item the most powerful peak with a median value of $\sim$45 days is near the expected time delay from the calibrations; 
\item the second and the less significant peak appears with a median value of $\sim$125 days.\\
\end{enumerate}
The expected value from the calibration relation for the object \#1 is $\sim$27~days, and we confidently take the position of a powerful peak  $\tau_1=$~\tone days as a preliminary result. \\
Clearly, with an increase in the number of epochs, we can reveal a narrowing of the main peak, thereby reducing the error in estimating the time delay value. 

\subsubsection{JAVELIN analysis for object \#2}
\begin{figure}
    \centering
    \includegraphics[width=0.48\textwidth]{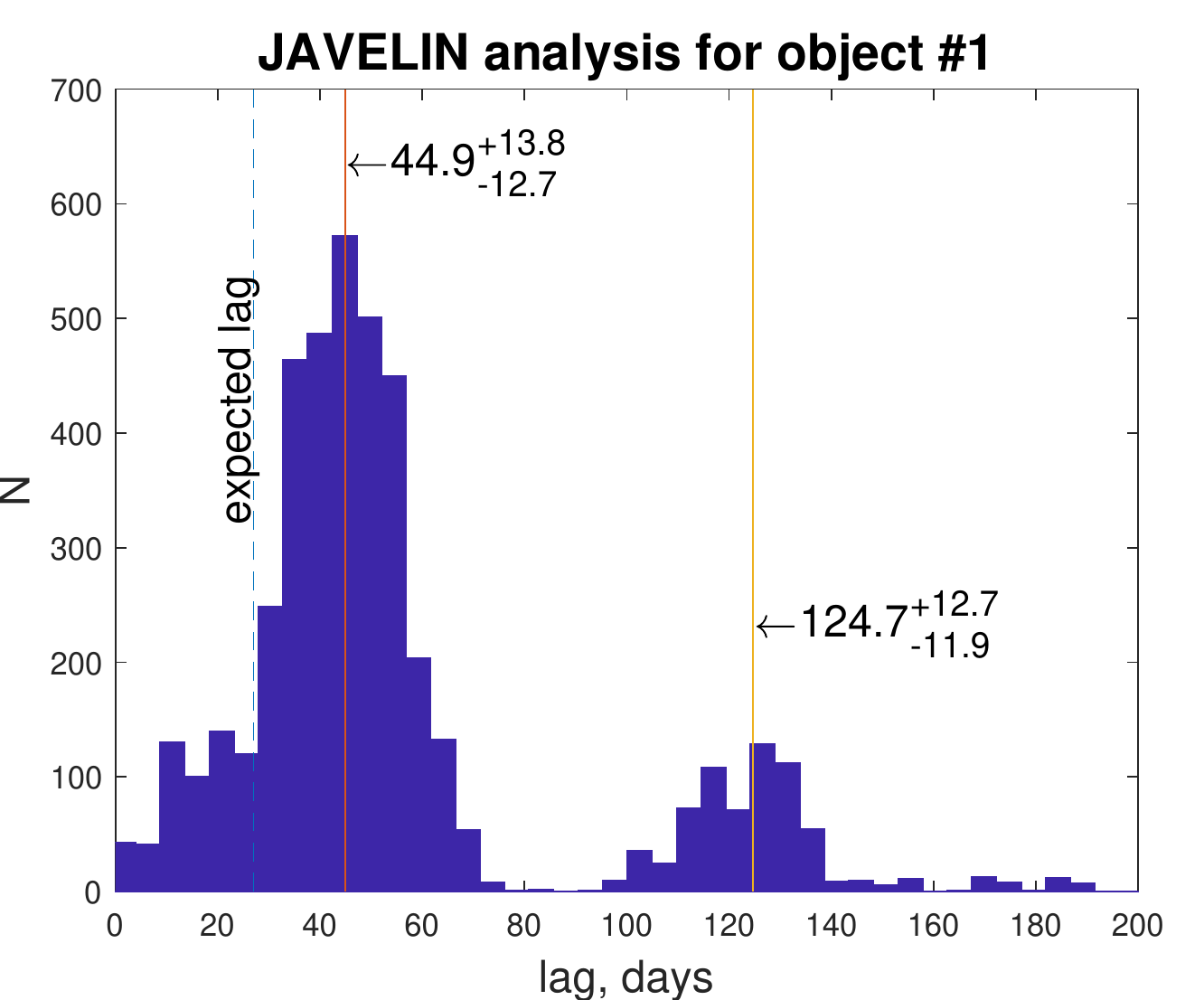}
    \includegraphics[width=0.48\textwidth]{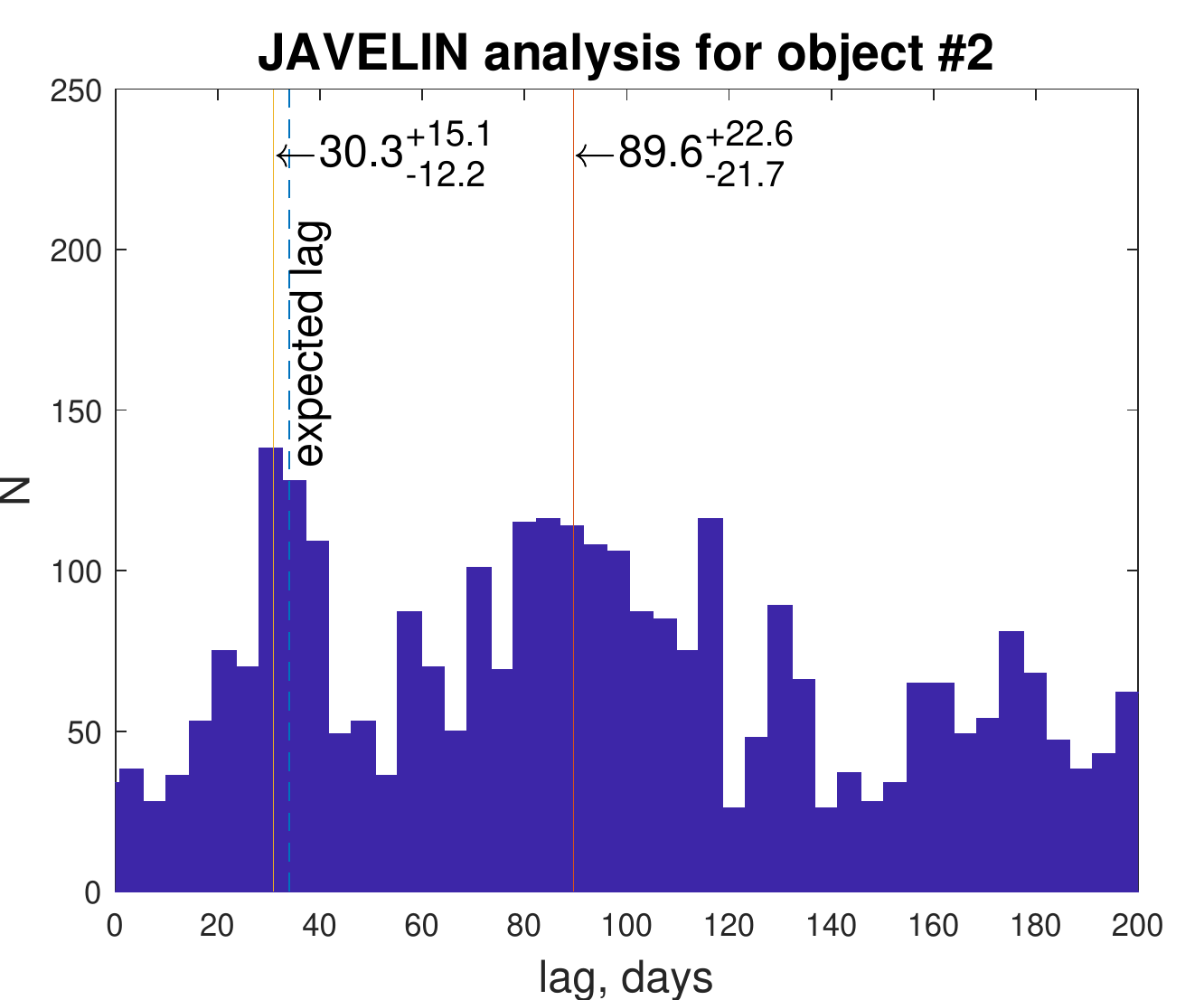}
    \caption{The time lag posterior distributions using {\tt JAVELIN} code for the object \#1 ({\it left}) and objects \#2 ({\it right}). Solid red lines indicate significant peaks and their values, and the expected values from calibrations are indicated by dashed blue lines.
    }
    \label{ccf}
\end{figure}

For the object \#2, the number of epochs is only 17, and we see a less clear picture (Fig.~\ref{ccf}, right panel). For the range of time delays from 0 to 150 days sampling results are divided into two main solutions with a comparable level of significance:

\begin{itemize}
    \item peak at $\sim$30 days, which is quite close to the value expected from the calibrations;
    \item wider peak with a median value of $\sim$90 days.
\end{itemize}

Despite the fact that we see two comparable peaks, we prefer the first peak, which is closer to the expected value $\sim$34~days from calibrations. An additional argument is that with an increase in the number of epochs we trace the growth of the first peak and hope to see this trend after additional observations. So, for the object \#2 we take $\tau=$~\ttwo days as a rough preliminary result. \\
The peaks are wide and still indistinguishable, so more data are needed for a more clear histogram.

\subsubsection{General comments}

We should not exclude the possibility of mismatch between the roughly expected time delays and the obtained peaks in the histogram, since the observed luminosity of AGN is variable. That is, the fainter the state of the nucleus we observe, the smaller BLR size estimate is. Moreover, the spectral data for objects \#1 and \#2 were obtained more than 20 years ago, so the new spectral observation should be done.
Also, certainly, the effect of sparse and poor sampling and low number of epochs influence the analysis of light curves. Obviously, we would increase the number of epochs and to clarify our results further we would also compare different cross-correlation methods.



\section{Conclusions}

In this paper we present the first results of the photometric reverberation mapping project started in 2018 on the 1-m telescope of SAO RAS. The ongoing project is focused on the regular observations of the sample of 8 broad line AGNs, and here the technique of observation and data reduction is shown and the preliminary results demonstrating the method stability.
Within this work, the following was obtained.

1. The observations by the photometric reverberation mapping method are adapted for telescopes of 1-meter class and are independent of the device used.

2. For each of the studied AGN in the range of redshifts $0.1 < z < 0.8$, a list of local standards was determined, which allows further use of the differential photometry method. The photometric accuracy is on average 0.03 mag, which is an order of magnitude greater than the expected amplitude of the AGN variability.

3. The use of the {\tt JAVELIN} analysis revealed time delays  $\tau_1 =$ \tone and $\tau_2 =$ \ttwo days for objects \#1 and \#2, respectively.

It is clear that the first obtained results are in agreement, within the error-bars, with our predictions based on the empirical radius-luminosity relations. However, the data should be processed more carefully and other cross-correlation methods should be applied to reveal whether the calculated time delays are correct. Also, it is particularly important to clean the series from the additional harmonics. Presently, for the majority of the objects the sampling of the observations and the number of epochs is still not sufficient for the reliable analysis, and this is probably the reason why the cross-correlation peaks are so wide. We continue with the observations and in the next stage of the project we are going to obtain and analyze more data, and compare different cross-correlation methods.


\section*{Acknowledgements}
The work is executed at support of RFBR grant 18-32-00826. The authors thank V. L. Afanasiev for useful discussions and comments.
Observations on the telescopes of SAO RAS are carried out with the support of the Ministry of science of the Russian Federation.
%
%
%
\bibliographystyle{mnras}
\bibliography{BLR}

\end{document}